\documentclass[twocolumn,amsmath,amssymb,pre,aps]{revtex4}   
\usepackage{graphicx}
\usepackage{dcolumn}
\usepackage{bm}
\usepackage{color}
\usepackage{graphicx,wrapfig,lipsum}
\usepackage{hyperref}
\usepackage{float}

\renewcommand{\vec}[1]{\mathbf{#1}}
\newcommand{\squeezeup}{\hspace{-40.0mm}}

\newif\ifgraph

\graphtrue             %

\begin{document}
\title{Escape Kinetics of an Underdamped Colloidal Particle from a Cavity through Narrow Pores}

\author{Shubhadip Nayak\textsuperscript{a}, Tanwi Debnath\textsuperscript{b}, Shovan Das\textsuperscript{a}, Debajyoti Debnath\textsuperscript{a} and Pulak K. Ghosh\textsuperscript{a}\footnote[1]{Email: pulak.chem@presiuniv.ac.in}}

\affiliation{\textit{$^{a}$~Department of Chemistry, Presidency University, Kolkata 700073, India}}
\affiliation{\textit{$^{b}$~Department of Chemistry, University of Calcutta, Kolkata 700009, India}}

\date{\today}

\begin{abstract}
\begin{wrapfigure}{r}{7.0cm}
\squeezeup
\includegraphics[width=7.0cm, height=4.0cm]{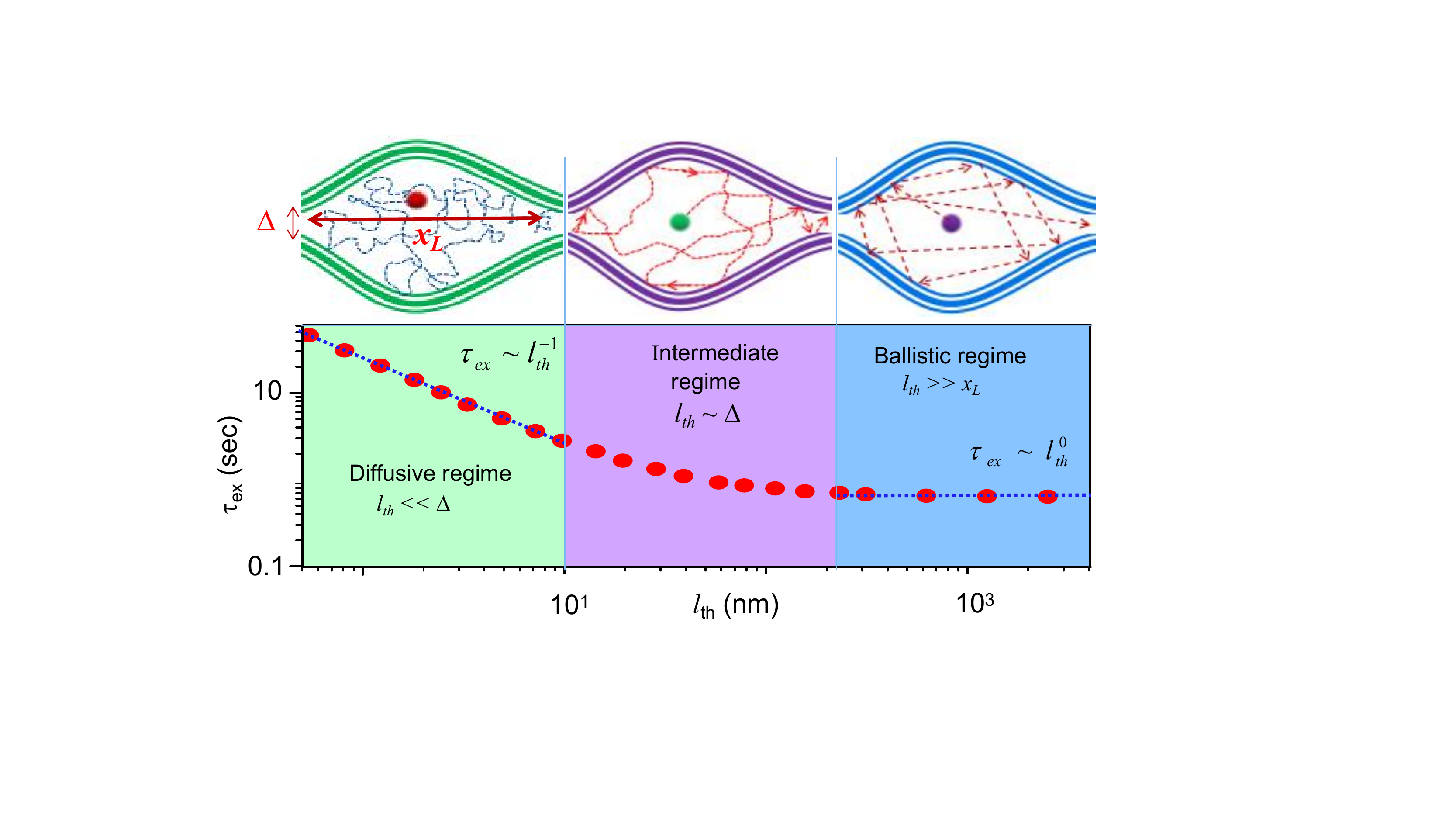}
\squeezeup
\end{wrapfigure}
  It is often desirable to know the controlling mechanism of survival probability of nano - or microscale particles in small cavities such as, e.g., confined submicron  particles in fiber beds of high-efficiency filter media or ions/small molecules in confined cellular structures. Here we address this issue based on numerical study of the escape kinetics of inertial Brownian colloidal particles from various types of cavities with single and multiple pores. We consider both the situations of strong and weak viscous damping. Our simulation results show that as long as the thermal length is larger than the cavity size the mean exit time remains insensitive to the medium viscous damping. On further increasing damping strength, a linear relation between  escape rate and damping strength emerges  gradually. This result is in sharp contrast to the energy barrier crossing dynamics where the escape rate exhibits a turnover behavior as a function of the damping strength.  Moreover, in the ballistic regime, the exit rate is directly  proportional to the pore width and the thermal velocity. All these attributes are insensitive to the cavity as well as the pore structures. Further, we show that the effects of pore structure variation on the escape kinetics are conspicuously different in the low damping regimes compared to the overdamped situation. Apart from direct applications in biology and nanotechnology, our simulation results can potentially be used to understand diffusion of living or artificial micro/nano objects, such as bacteria, virus, Janus Particle etc. where memory effects play dictating roles.  
 
  
\end{abstract}


\maketitle

\section{Introduction}\label{intro}
\begin{figure}[ht]
\centering \includegraphics[width=6cm]{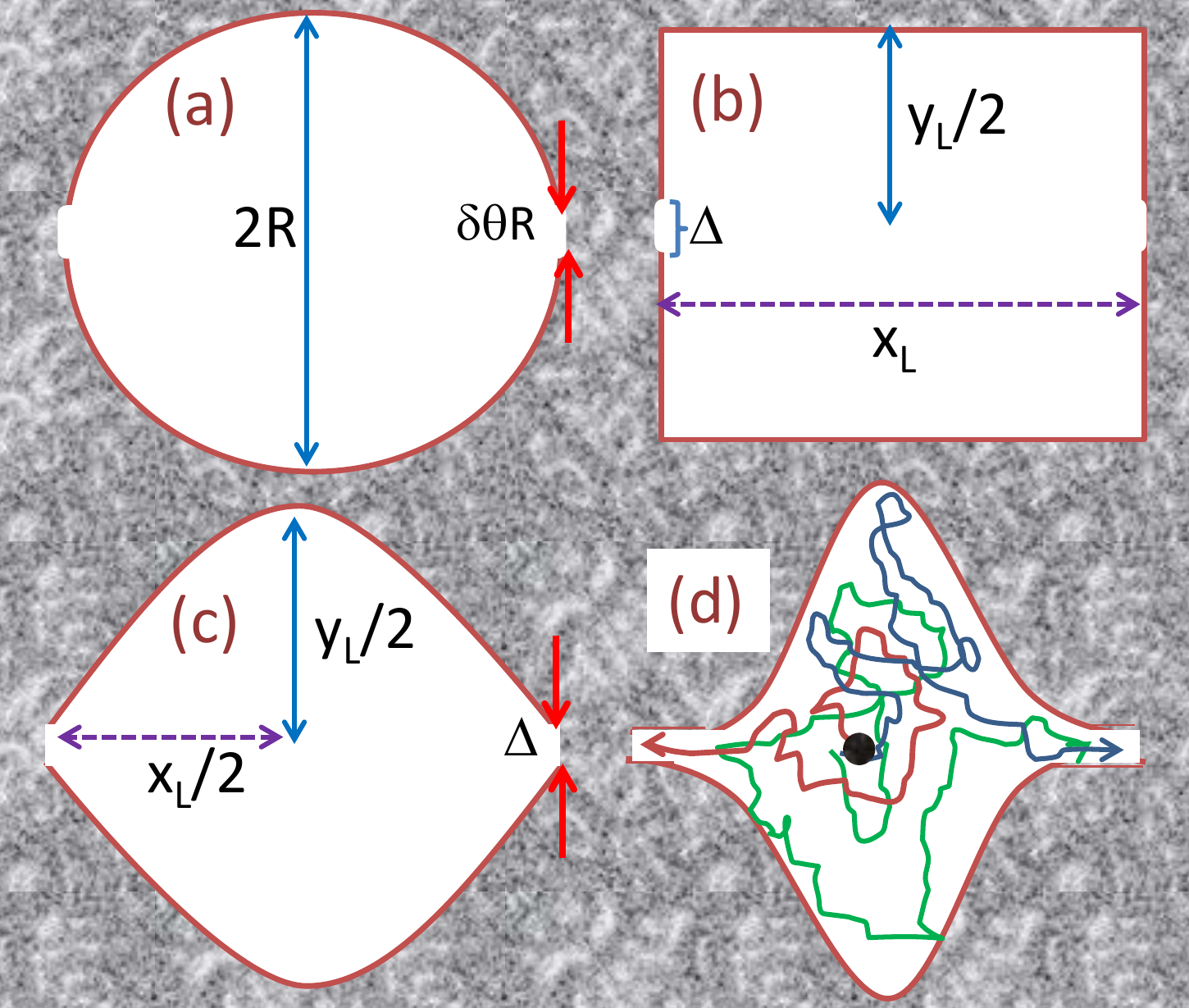}
\caption{ (Color online) Schematic of various types of cavities; (a) Circular cavity of radius $R$ with two narrow openings of equal size $R \delta \theta$. The centers of two openings are antipodal to each other. (b) Rectangular box ($x_L \times y_L$) with two openings of the equal width $\Delta$.  (c) and (d) represent compartments of  smoothly corrugated channels. The channel cross-section has been modeled by a periodic function $\pm w(x)$ (see Eq(\ref{corrugated-walls})). The channel width maximum, minimum and the compartment length are $y_L$, $\Delta$ and $x_L$, respectively. Here, $\eta$ determines the openings as well as  the cavity shapes:  the pore assumes a triangular shape when $\eta = 1$ (c);  the pores take capillary like shape $\eta = 4$ (d). \label{F1}}
\end{figure}
Diffusion is the ubiquitous mechanism for intercellular transport of key bio-molecules that trigger many cellular functions~\cite{Bio1,Bio2,Bio3}. Examples include cross-membrane ions or molecules transfer to maintain ionic or chemical balances, flowing in (out) nutrients (waste) into (from) the bloodstream, electrical signals moving down the length of a neuron's arm  etc. to name a few.  All these processes involve diffusive transport of particles through narrow openings of membrane walls~\cite{Bio1,Bio2}. Thus, the escape processes actually play the central role in governing the timescales of many cellular functions. Moreover, all these processes are associated with some sort of entropic barrier crossing events dictated by the geometry of cellular structures in addition to other factors (like,  density gradients, temperature, solvation , pH, interaction with substrate   etc). Therefore, it would be desirable to  have  knowledge about the timescale and controlling mechanisms of these key cellular events ~{\cite{Bio3,Hangi-rev2}}. It could allow one to rule over cellular function by tuning external parameters.

Further, working mechanisms of many nano-devices rest on the diffusion of  Brownian traces in suitably tailored confined structures. For example, the particle capturing mechanism on the fibers of the high efficiency filter media. The fidelity of capturing processes here depends on the particles residence time inside the small cavities in the internal micro-structures  of the filters. Moreover, depending upon their size particles get absorbed on the fibers through inertial impaction or interception or diffusion mechanism. To this end, a number of studies focus on escape kinetics of Brownian particles from cavities of various structures that can potentially mimic diffusion controlled cellular processes to a great extent. Particularly, some earlier investigations ~\cite{holcman-review, jcp1, jcp2}  on  escape kinetics from circular cavities and corrugated channel compartments certainly provide suggestive opinions.

Recognizing the importance of  diffusion under geometric constraints  in nanotechnology and biology, its various facets have been explored \cite{R-PRL1,R-physics-report,R-PRL2,R-JACS,hanggi-review,Burada-review,HangFJ1,HangFJ2}.   Diffusion through narrow tubes with varying  cross-section has been investigated in a number of contexts related to driven transport through artificial nanopores ~\cite{ Burada-review,HangFJ1,HangFJ2,hanggi-review,JPCC-theo,JPCC-interface,Bao-1,JPCC-zeolites,Bao-2} or  biological channels ~\cite{Kullman,JPCC-oxygen-diff}. The entropic variants of noise-induced phenomena ~\cite{ourSR1,ourSR2,DSR1,DSR2, HangSR}  ( like, stochastic resonance, resonant activation, noise-induced asymmetric localization, ratcheting etc.) have been explored to realise constructive roles of fluctuations when uneven boundary effects govern diffusion. Some recent studies focus on transport properties of artificial microswimmer~\cite{JP1,JP2,Bao-3,Bao-4,JPCC-Janus1,JPCC-Janus2,Lowen-JP,volpe-1} (e.g., Janus particles) in confined  structures aiming at emerging applications of active particles in nanotechnology and medical sciences.

Most of the previous studies on confined diffusion focus on the highly damped situation where inertia practically plays no role and thermal fluctuations are assumed to be white. This assumption is valid when viscous relaxation time ($\tau_{\gamma}$) is orders of magnitude shorter than any other relevant  timescale of the system.  However,  in the short time regime (($t <\tau_{\gamma}$)),  neither inertia effects nor corrections in thermal fluctuations can safely be ignored.  Recent developments in optical trapping and advanced detection methods make it possible to experimentally access dynamical features of Brownian particles in the short time regimes~\cite{Raizen,nature-exp1,nature-exp2,Lowen-inertia1,Lowen-inertia2,Raizen2}.  For diffusion in confined structures, the validity of the assumption of the overdamped limit depends on the size of confining cavity as well as the relative amplitude of the particle mass, viscous damping and average thermal velocity~\cite{Our-inertia}. 
  
In this paper we explore the escape kinetics of an underdamped Brownian colloidal particle from different types of cavity with varying pore structures.  To be specific, we consider three types of 2D confining cavities with narrow openings: (a) Circular cavity, (b) rectangular cavity and (c) compartment of a smoothly corrugated channel. 
They are depicted in the Fig.~1. The particle dynamics in the confined geometries are largely guided by uneven boundary effects as well as  the fluctuation statistics. Again, the boundary effects are determined by the particles shape and size as well as the structure of the confining walls. For the sake of simplicity, in the present study, we consider structure less point-size particles. Thus, the boundary effects in the escape process are solely dictated by the wall structures. Moreover, in the low damping regime particles undergo ballistic motion, thereby, the boundary  effects become apparently different from the overdamped  particles exhibiting uncorrelated motion.  

Our simulation results show that in sharp contrast to the overdamped limit, underdamped particles escape rate is proportional to the pore width  and the square root of the temperature irrespective of the pore structures.   Moreover, effects of pore structures on the escape kinetics are noticeably different in the three regimes of viscous damping. For example, consider the change of pore tips  from cusp-like to sinusoidally convex through a triangular shape [as shown in Fig.~4(c) (inset)]. In this sequence of the structural variation, the escape rate very slowly enhances in the low damping regime and almost remains unchanged for intermediate damping. However, in the overdamped limit, the escape rate gradually suppresses.

The outlay of the paper is as follows. In Sec.~\ref{Model} we present  Langevin model describing dynamics for inertial Brownian particles in 2D.  In Sec.~\ref{Results}, we present our simulation results for escape kinetics from various types of cavity. We first consider the circular cavity with single and multiple pores, then, the variation of cavity as well as the pore structures. In Sec.~\ref{Conclusions} we summarize important results with concluding remarks.

\section{Model}\label{Model} We consider a point-like Brownian particle of mass
$m$ diffusing in a 2D suspension fluid contained in a  cavity.  The particle can escape from the cavity through  narrow openings  as illustrated in Fig.~ 1.  The dynamics of the particle (in the xy-plane) is modeled by
the 2D Langevin equations, 
\begin{eqnarray}\label{Langevin}
m\ddot{q}= -\gamma \dot{q} + \sqrt{2 \gamma k_BT}\xi_q(t)
\end{eqnarray}

where, $q = x, y$ and $\{x, y\}$ is the instantaneous position of the particle.  The random forces, $\{\xi_x(t), \xi_y(t)\}$
are modeled by Gaussian noises with zero-mean, $\langle \xi_i \rangle=0$ and delta correction, $\langle \xi_i(t')\xi_j (t) \rangle=\delta(t-t')$, where $i,j = x, y$.  For a spherical particle of radius $r_0$ moving in a medium with viscosity $\eta_v$, the damping constant $\gamma$ can be estimated (based on Stokes relation) as, $\gamma = 6\pi \eta_v r_0$. However, in the present context we assume the damping constant plays the role of an effective viscous force incorporating all additional effects that are not explicitly accounted for in Eq.~ (\ref{Langevin}), like hydrodynamic drag, particle-wall interactions, etc.  For a free Brownian particle, the mean square displacement (MSD) grows with time as, $\langle\Delta x^2 \rangle= \langle\Delta y^2 \rangle = 2Dt+2D \tau_\gamma (\exp[-t/\tau_\gamma]-1)$. Where, the diffusion constant, $D=k_B T/\gamma$  and the viscous relaxation time, $\tau_{\gamma} = m/\gamma$.  At short times $t << \tau_\gamma$, the MSD  exhibits quadratic growth in time, while at long times $t >> \tau_\gamma$, the MSD is linear in time. The dynamical nature of the particle, whether ballistic or diffusive, is determined by the cavity size relative to the thermal length.

It is a formidable task to  solve the Eq.~(\ref{Langevin}) analytically with non-holonomic constraints. Also, probabilistic Fokker-Planck description based on the so called Ficks-Jacobs approximation is limited to smooth confining structures and  highly viscous medium~\cite{HangFJ1,HangFJ2,jcp2}.  This difficulty has been circumvented by numerically solving the  Eq(\ref{Langevin}) using a standard Milstein algorithm. The numerical integrations have been performed using a very short time step, $10^{-4}-10^{-5}$, to ensure numerical stability. We impose elastic reflection of velocity $\dot{r}$ at the boundary walls. 

 To be specific, we simulate dynamics of particles' center of mass by considering point size particles in a cavity. Thus, in our model, cavity volume is the free space accessible to the center of mass.  Numerical values of cavity width and the bottle neck  (pore size) reported in simulation parameters are actually meant for the center of mass. The cavity space accessible to the particles center of mass and the corresponding effective cavity volume are schematically shown in Figure S1 ( see supporting information).

 We estimate the mean exit time $\tau_{ex}$ out of various types of cavities.  Here, $\tau_{ex}$ is defined as the average time the particle takes to exit the cavity through a very narrow exit window. Initially, the particle is placed at the center of the cavity with a random velocity.  Further,  Maxwellian distribution of initial velocities has been assumed. The simulation results reported in  this paper were obtained as ensemble averages over $10^6$ trajectories. 

For the simulation parameters values used here,  length and time scales are  micrometers and seconds, respectively. We consider the mass of the particle is the unit of mass. The unit mass can be estimated as follows. Let us consider  a slilica bead of radius 1$\mu m$. Taking the density~\cite{mass} of SiO$_2$ 2 g/cm$^3$, the  mass of the particle would be
about $8.3 \times 10^{-12}$ g. 

\section{Results and discussions}\label{Results}
Mean exit time $\tau_{ex}$ for various cavity structures as a function of damping (thermal length), pore sizes and other parameters are presented in the Fig.~2 to Fig.~4.  We characterize colloidal particles  by two parameters, thermal length and thermal velocity. In our simulation, the thermal velocity is of the order of $0.1 \mu m/s$ [except in Fig.2(a,b)]. We show in the supporting information (page 2-3, items A2- A4) that changing velocity amounts to re-scaling time. Thus, none of the key features of escape kinetics can be changed by tuning velocity for given thermal lengths.  

  We mainly stress on the low to intermediate damping regimes. The results for large damping (also reported elsewhere) set references for our analysis.  Based on the relative amplitude of the thermal length, $l_{th}$ = $(m/\gamma)\sqrt{k_B T/m}={\rm v_{th}\tau_{\gamma}}$ to the other length scales of the system, our analysis divides damping strength into three regimes: low, intermediate and high. In our simulations, the thermal length is varied by changing damping strength  $\gamma$ for a given thermal velocity  ${\rm v_{th}}$. Thus, in our discussion $\tau_{ex}$  {\it vs.} $l_{th}$ is equivalent to $\tau_{ex}$  {\it vs.} $\gamma$.

 We first systematically consider the escape kinetics of an underdamped particle out of circular cavities with single, double  and multiple openings. Next, to analyse the effect of cavity and pore structures we study exit rate from rectangular cavities and compartments of corrugated channel with various pore geometries.
\subsection{Circular cavity}
 \noindent { \it Cavity with a single exit window} --- Let us begin with escape kinetics from a circular cavity of radius $R$ with a single opening of arch length $\Delta = R \delta \theta$. The particle can escape only through the polar angle $\delta \theta$. Figure 2(a) and  2(b) depict variations of the mean exit time $\tau_{ex}$ as a function of the pore width $\Delta$ and the thermal length $l_{th}$, respectively. Three types of plot in  $\tau_{ex}$ versus $\Delta$ can easily be recognized. They correspond to the two opposite limits, $ l_{th} >> R$ and $ l_{th} << \Delta$, and the intermediate one $ l_{th} \sim \Delta$. For the first case, the mean escape time is insensitive to the medium viscosity  and  inversely related to  the opening size.  On the other hand, in the highly viscous medium, $\Delta >> l_{th}$, the mean exit time exhibits very slow logarithmic decay with increasing pore width and  linear relation to the damping constant. For the intermediate values of damping strength, $ l_{th} \sim \Delta$, the exit time first follows an inverse relation with $\Delta$, then exhibits a logarithmic decay.   Similar features are manifested in $\tau_{ex}$ versus $l_{th}$ through the appearance of  two distinct segments, the plateau and the decaying branch [see Fig.~2(b)]. Two segments are well separated by a large window of $l_{th}\;(\equiv\gamma)$ value. We now detail features of escape kinetics in the three regimes of damping separately.
 %
\begin{figure}[tp]
\centering \includegraphics[width=7.0cm]{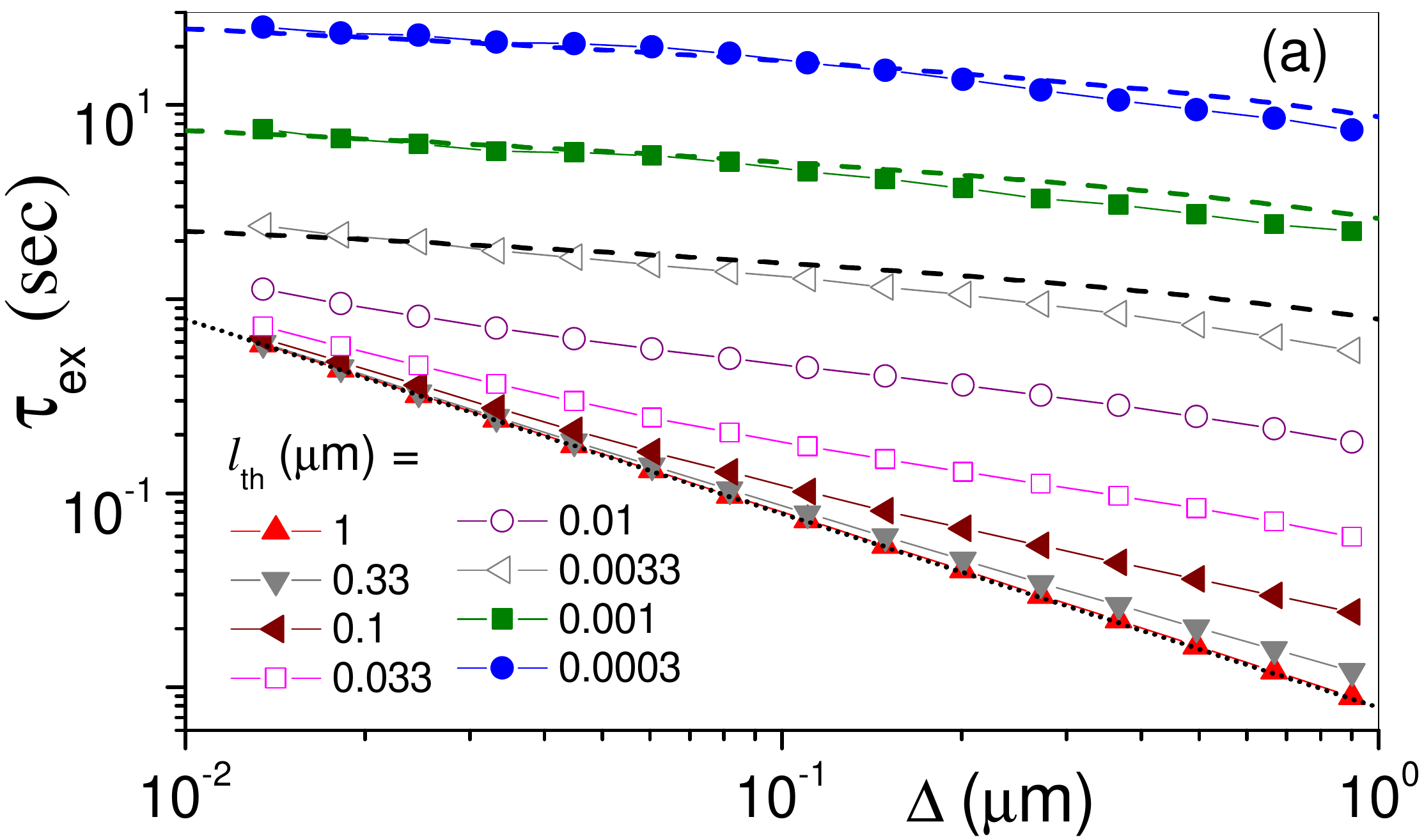}
\centering \includegraphics[width=7.0cm]{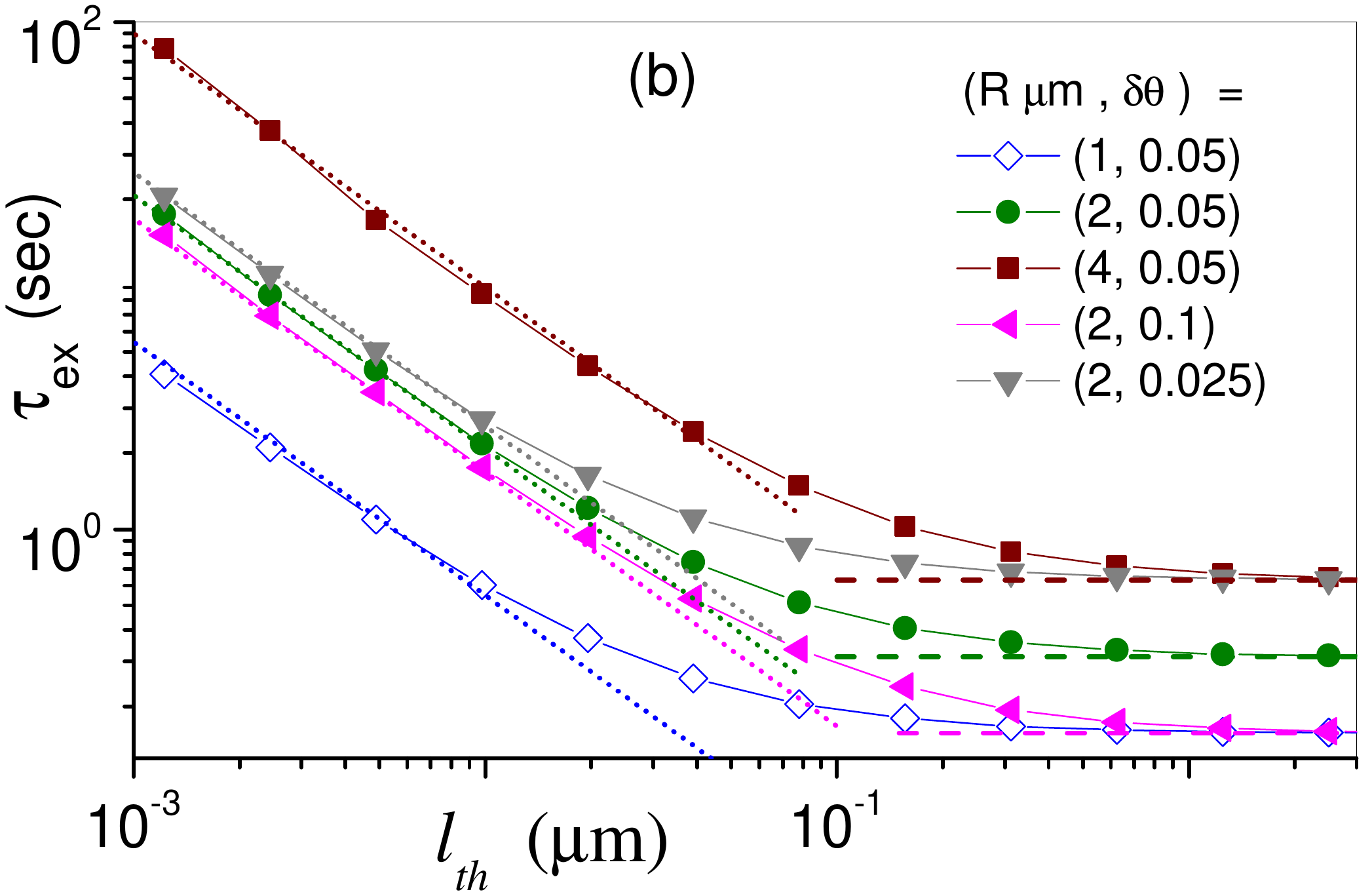}
\caption{(a) Mean exit time $\tau_{ex}$ from a circular cavity as a function of $\Delta  $ for various thermal length $l_{th}$. Dotted line is the analytical estimate of $\tau_{ex}$ for the very low damping regime using Eq.~(\ref{MET-UD}). Dashed line is the analytical prediction for the overdamped limit based on the Eq.~(\ref{MET-OD}). (b) Mean exit time $\tau_{ex}$ versus thermal length $l_{th}$ for different pore size, $\delta \theta$ and the cavity radius $R$. Dashed lines and dotted lines represent the analytical estimates based on Eq.~(\ref{MET-UD}) and Eq.~(\ref{MET-OD}), respectively. Simulation parameters (unless mentioned in the legends): ${\rm v_{th}} = 1000\;\mu m/s,\; R = 1\;\mu m,$ and $\delta \theta = 0.1$.  \label{F2}}
\end{figure}

 (a)   The low damping region is identified by the condition,  $ l_{th} >> R $. This situation corresponds to quite  a large viscous relaxation time  $\tau_{\gamma}$,  so that the distance travel by the particle  with thermal velocity $\sqrt{k_BT/m}$ within the time $\tau_{\gamma}$ is much larger than the confinement size. Thus, the particle moves back and forth inside the cavity and occasionally exits through the narrow opening with a probability  $\delta \theta/2\pi$. If we set the particle at the center of the cavity with velocity $\vec v_0$, exit at the first attempt will take the time, $\tau_1 = R/v_0$. The exit time for 2nd, 3rd .... attempts would be, $\tau_2 = 3R/v_0,\; \tau_3 = 5R/v_0$ .... , respectively. On an average, the confined particle exits the circular cavity within $\bar{n} = 2\pi/\delta \theta$ attempts. Thus, the mean exit time can be estimated as,
\begin{eqnarray} \label{MET-UD}
\tau_{ex} \simeq \frac{1}{\bar{n}}\sum_{i=1}^{\bar{n}} \tau_i = \frac{2 \pi R}{\delta \theta} \sqrt{\frac{\pi m}{2k_B T}}.
\end{eqnarray}
This estimate rests on the two assumptions: (i) Viscous relaxation time is so large that particles do not change their direction until they collide the wall. (ii) Initial velocity has Maxwellian distribution, so the particle drifts to the boundary wall with an average velocity, $\bar{v} = \sqrt{2k_B T/\pi m}$ . Our numerical simulation results in Fig.~2(a,b)  completely agree with Eq.~(\ref{MET-UD}) as long as the thermal length is larger than the diameter of the cavity.  

 (b) In the overdamped limit, both the thermal length and the viscous relaxation time are much shorter than  any other relevant length and time scales of the system, respectively. In this regime, the exit rate and other transport quantifiers, like, diffusivity and mobility are inversely related to the damping strength $\gamma$. Thus, one can easily scale out $\gamma$ from  all these observables. The mean exit time in this limit for very narrow opening is given by~\cite{holcman-review} 
\begin{eqnarray} \label{MET-OD}
\tau_{ex} \simeq \frac{\gamma R^2}{k_B T} \left\{ \ln\left[\frac{2\left( 2\pi - \delta \theta \right)}{\delta \theta}\right]+\frac{1}{4}\right\}.
\end{eqnarray}
 Figure 2(a,b)  shows that simulation results in the overdamped limit are in good agreement with the analytic estimate for very small $\Delta$. However, the deviation becomes noticeable with increasing the size of exit window as  corrections due to   terms $O(2\pi/\delta \theta$) and higher orders become relevant \cite{holcman-review}. In the sharp contrast to the low damping limit, here mean exit time is proportional to the cavity volume, damping constant and has a slow logarithmic dependence on the opening size $\Delta$. All these features are apparent in Fig.~2(a,b).
 
  (c) In between two extreme regimes, there is a large window of damping strength where ballistic motion persists over a length shorter than the cavity size. At the same time, the effect of finite correction in the displacement due to inertia cannot be discarded safely.  Here, neither Eq(\ref{MET-UD}) nor  Eq(\ref{MET-OD}) alone can estimate the mean exit time. Summing up the  contribution from both the extreme limits, the mean escape time can be expressed as, $\tau_{ex} \simeq \tau_{ex}^{ov}+\tau_{ex}^{ud}$. Where, $ \tau_{ex}^{ov}$ and $\tau_{ex}^{ud}$ are the exit time in the overdamped and underdamped limits, receptively. This estimation is valid for the entire range of viscous damping. One can easily identify intermediate regimes from the plots in Fig.~2(b). It is delimited between the onset of deviation from Eq.~(\ref{MET-UD}) and the starting point of the linear regime. Our simulation results show that the intermediate regime is extended over almost two orders of magnitude.   Further, simulation results  in Figure 2(a,b) show that depending upon the pore size, the starting point of deviation from the overdamped regime falls within the thermal length 0.01 $\mu m$ to 0.05 $\mu m$. To be specific, consider the data set [in Fig 2(b)] for $\delta \theta = 0.1$ ($\Delta = 0.2 \;\mu m$) and $R  = 2\;\mu m $. The inertia effect begins noticeable at $l_{th} \sim 25 \; nm$.  Again, the thermal length of micron size particles at room temperature fall in the range 0.1 nm to 50 nm depending upon the medium viscosity and mass density of the colloids. Thus, for diffusion through very narrow porous structures, in many practical situations, one cannot safely ignore inertial effects. The details about experimental accessibility of the inertial impact shown in the supporting information (item A5, and figure S3).    

\noindent {\it Cavity with double exit windows} --- We now examine the effects of an additional exit window on the mean exit time from a circular cavity. We consider pore centres are separated  by an angle $\theta$ as shown in the inset sketch of Fig.~3(a). Therefore, the cordial distance between two opening centres is, $L = 2R\sin(\theta/2)$. Figure 3(a) represents mean exit time $\tau_{ex}$ as a function of $\theta$ for different damping  strengths/thermal lengths. When pores are well separated, exit through one window is not affected by the others. Thus, total exit rate can be expressed as a sum of the exit rate through individual windows. This leads to, $1/\tau_{ex} = \sum_i 1/\tilde{\tau}_{i}$. Where, $\tilde{\tau}_{i}$ is the mean exit time from the $i$-th exit window. For two exit windows of equal size, one can write, $\tau_{ex} = \tilde{\tau}/2$. Where,  $\tilde{\tau}$ is the mean exit time through one opening when the other window is closed. The Eq.~(\ref{MET-UD}) and Eq.~(\ref{MET-OD}) can be used  to obtain $\tilde{\tau}$ in the low and high damping limits, respectively. 

To better analyse the effects of separation between exit windows on the escape kinetics, we choose the exit ratio, defined as,  $\tau_R = \tau_{ex}(\theta)/\tau_{ex}(\pi)$, as a quantifier.  Here, $\tau_{ex}(\pi)$ is the mean exit time when the two openings are antipodal to each other. As the arch length of each exit window is $R\delta \theta$, two holes are merged at $\theta = 2\delta \theta$.

\begin{figure}[H]
\centering \includegraphics[width=7.5cm]{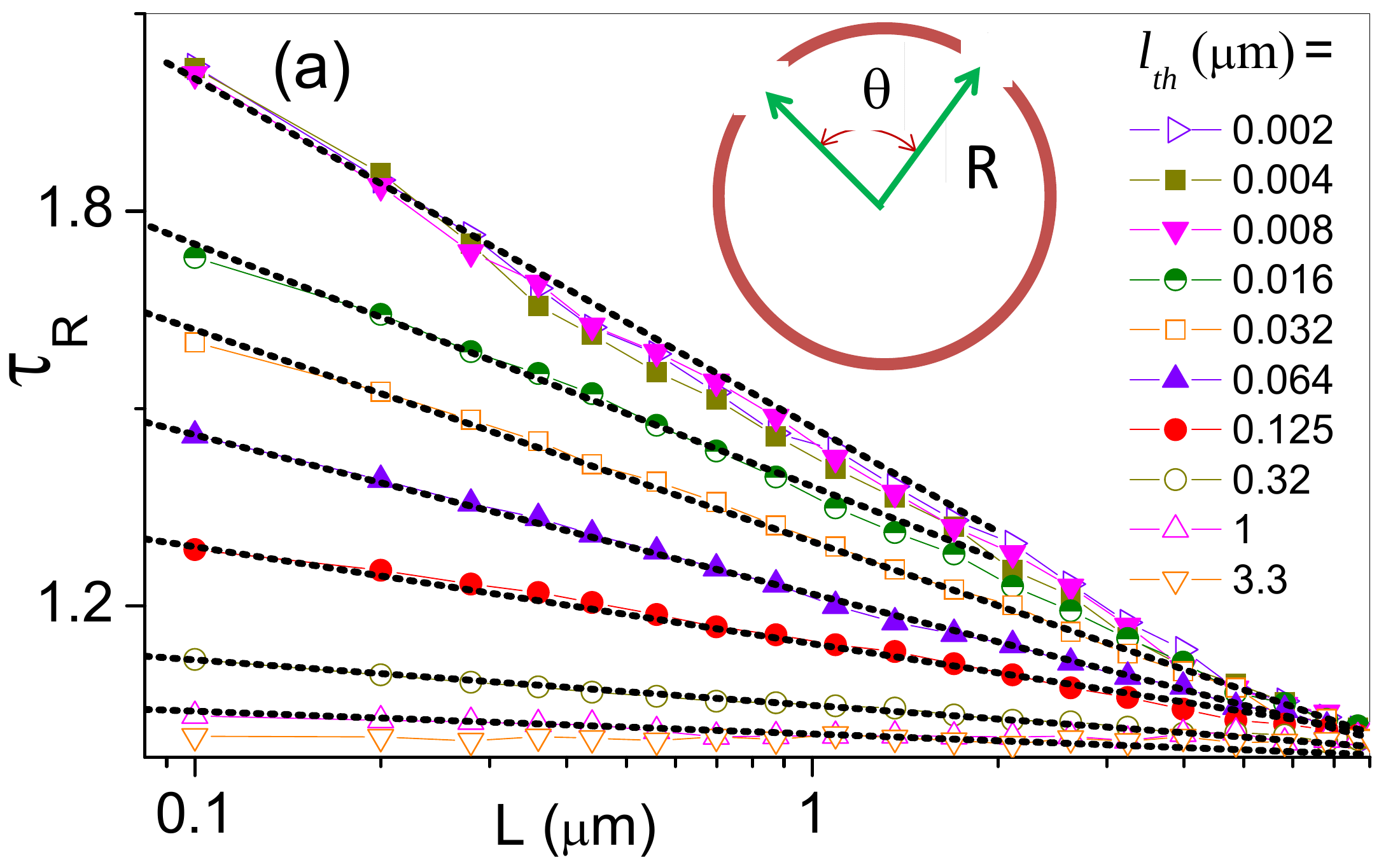}
\centering \includegraphics[width=7.5cm]{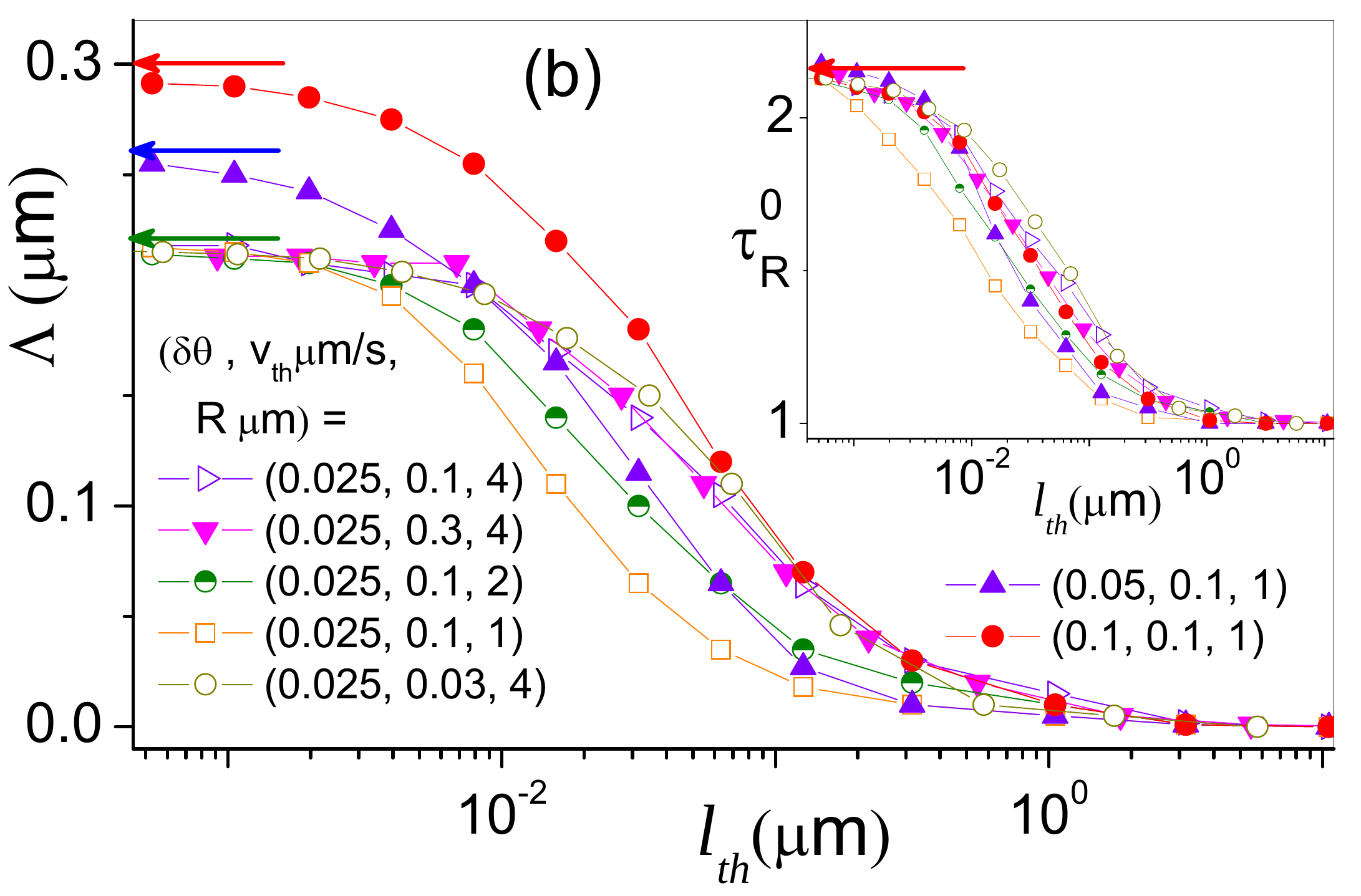}
\centering \includegraphics[width=7.5cm]{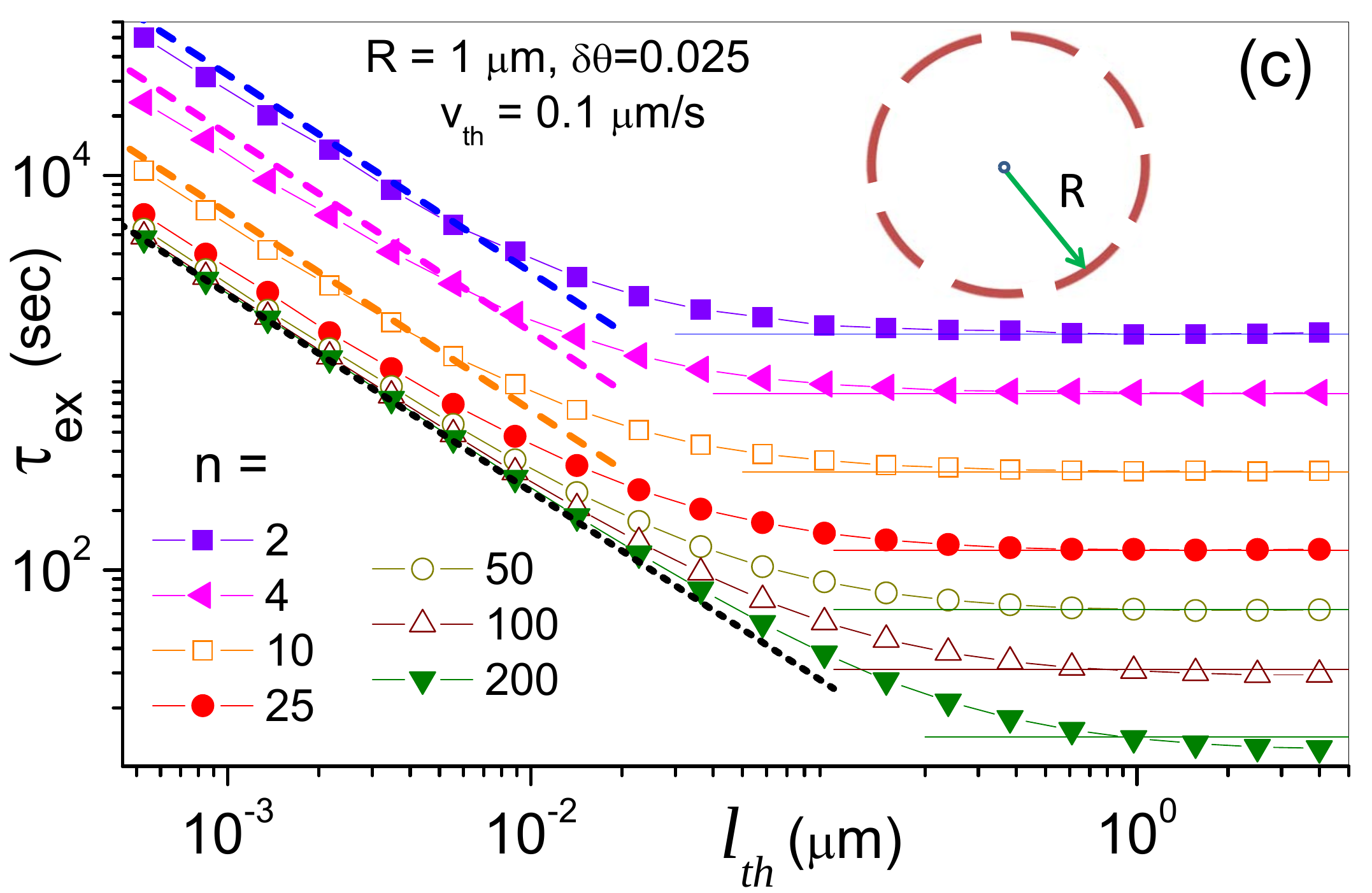}
\caption{(a) Exit time ratio, $\tau_R$ {\it vs.} cordial distance $L$  for various  $l_{th}$ values.  Dotted line is the analytical estimate of $\tau_R$ with best fitted slope $\Lambda$ and intercept $\tau_R^0$. Simulation parameters: ${ \rm v_{th}} = 0.32 \;\mu m/s, \; \delta \theta = 0.025, \; R = 4 \;\mu m$. (b) Slope $\Lambda$ [see Eq.~(\ref{ratio-emp})] {\it vs.}  $l_{th}$ for different opening width, thermal velocity and cavity size as mentioned in the legends.   Inset: The intercept $\tau_R^0$ {\it vs.} $l_{th}$. Both the slope and intercept are extracted based on the least square fitting. The upper bounds for the slopes and intercept are based on the Eq.~(\ref{ratio-asy}) and Eq.~(\ref{ratio-max}), respectively, indicated by horizontal arrows.  (c)  $\tau_{ex}$ {\it vs.} $l_{th}$ for different number of openings.  Horizontal solid lines are the analytical results for low damping limits based on the Eq.~(\ref{MET-multipore-UD}).  The dashed lines represent the analytical estimates for well separated pores in the overdamped limit [$\tau_{ex}=\tilde{\tau}_{exit}/n$ (see text)]. The dotted line represents Eq.~(\ref{MET-multipore-OD}).   \label{F3}}
\end{figure} 

 Thus, $\theta$ can be varied in between, $2\delta \theta $ to $\pi$. Figure 3(a) depicts the exit time ratio $\tau_R$ as a function of the cordial distance $L$ between two pores. It is apparent from this figure, in the very low damping regime the exit time ratio is insensitive to $L$ as well as the damping constant $\gamma \;(\equiv l_{th})$. In this regime, the particle  exhibits ballistic diffusion within the cavity as the thermal length $l_{th}$ larger than the cavity diameter $2R$. Here, exit probability is solely governed by the opening size but not on their relative positions. Thus, $\tau_R = 1$ for the entire $L$ range.  

With increasing damping strength the exit time ratio $\tau_{R}$ grows sensitive to the separating angle between two pores. It exhibits  a logarithmic decay as a function of $L$ with a $\gamma$ dependent prefactor. However, in the overdamped limit the prefactor becomes independent of  the damping strength and the exit time ratio reaches to its upper bound,  
\begin{eqnarray} \label{ratio-max}
\tau_R^{max} = \frac{2\left[ \ln\left({4\pi}/{\delta \theta}\right)+1/4\right]}{\ln\left({2\pi}/{\delta \theta}\right)+1/4}.
\end{eqnarray}
This estimate for different opening window sizes has been indicated by horizontal arrows in the Fig.~3(b) [inset]. The logarithmic decay of the exit time ratio $\tau_R$ with cordial distance $L$  can be expressed by the following empirical relation,
\begin{eqnarray} \label{ratio-emp}
\tau_R = \tau_R^0 - \Lambda \ln\left[ L/L_0\right]. 
\end{eqnarray}
Where, $L_0 $ and $\tau_R^0$ are the distance between the pore centers and the exit time ratio, respectively, for merging opening windows. Figure 3(a) shows that  simulation results are fitted well with the empirical relation (\ref{ratio-emp}) over the entire range of the damping constant $\gamma$. However, both the slope  $\Lambda$ and  intercept $\tau_R^0$ in $\tau_R$ versus $\ln\left[ L/L_0\right]$ are functions of the thermal length $l_{th}$. Figure 3(b) depicts variation of $\Lambda$ (main panel) and $\tau_R^0$ (inset) with $l_{th}$ for different thermal velocity, cavity and pore sizes. As mentioned earlier, we tune $l_{th}$ by changing $\gamma$ for a fixed ${\rm v_{th}}$. The slope $\Lambda$ remains  zero as long as  $l_{th} >> 2R$ . With increasing $\gamma$, as soon as thermal length becomes smaller than the cavity size, $\Lambda$ linearly grows with $\gamma$. Upon further increasing damping strength, $\Lambda$ approaches to a saturation limit. Following Holcman {\it et. al.},~\cite{holcman-review} the asymptote can be approximated as,  
 \begin{eqnarray} \label{ratio-asy}
\Lambda = \left[\ln({2\pi}/{\delta \theta})+c\right]^{-1}.
\end{eqnarray}
where $c$ is a constant can be extracted from numerical simulation results. Both, the simulation and analytical estimates (\ref{ratio-asy}) show that the asymptotic value of $\Lambda$ is determined by the size of the exit window and is insensitive to the  volume of the confining cavity and the temperature. The intercept $\tau_R^0$ has similar $l_{th}$ dependence as  $\Lambda$ [see inset of Fig.~3(b)]. However, it is bound in between 1 to $\tau_R^{max}$.  Following the above discussion, it is apparent that both $\Lambda$ and $\tau_R^0$ are insensitive to $l_{th}$ as long as inertia effects can be ignored. Thus, inspecting simulation data in Fig 3(a) one can immediately recognize that $l_{th} = 16 \; nm$ corresponds to inertial regime.

\noindent {\it Cavity with multiple exit windows} --- Figure 3(c) represents variation of the mean exit time with  thermal length when the confined particle has the option to choose one of the many narrow openings to exit.  Our study assumes that the exit pores are uniformly distributed over the circumference of the circular confinement as shown in the Fig.~3(c) [schematic in the inset]. In the low damping limit, due to ballistic motion the particle  drifts to the interior after colliding at the confining walls. Thus, the exit rate is affected by the adjacent pores no matter how close they are. Following Eq.~(\ref{MET-UD}) mean exit time is given by,
\begin{eqnarray} \label{MET-multipore-UD}
\tau_{ex} = \frac{2 \pi R}{n\delta \theta} \sqrt{\frac{\pi m}{2k_B T}.}
\end{eqnarray}
Where, $n$ is the number of exit windows. The estimates based on the above equation well collaborate simulation results in Fig.~3(c). On the other hand, in the highly damped situation ($l_{th} << \Delta$), as long as pores are well separated, the exit time can be estimated as, $\tau_{ex} = \tilde{\tau}_{exit}/n$. As noted earlier, $\tilde{\tau}_{exit}$ is the exit time from the cavity with a single opening window of width $R\delta \theta$. With decreasing spacing between pores, some sort of interference effect becomes operational and the exit rate exhibits a slow logarithmic growth with increasing pore number. For quite a large pore density, the mean exit time is simply equal to the diffusion time over the length $R$,
\begin{eqnarray} \label{MET-multipore-OD}
\tau_{ex} = {\gamma R^2}/{4k_B T} 
\end{eqnarray}  
However, such a limit of continuum cannot be realized for the particle with inertia. Due to memory effect, after colliding against  the wall particles are directed to the interior instead of diffusing to the adjacent exit points.

\subsection{Variation of cavity and pore structures}
We now examine how the escape kinetics of an underdamped Brownian particle is affected by the structural variation of the confining cavity as well as the exit window. To this purpose we consider escape from a compartment of  corrugated channels with wall profile function, 
\begin{eqnarray} \label{corrugated-walls}
w_{\pm} (x)= \pm (1/2)\left[\Delta +(y_L - \Delta)\sin^{\eta}(\pi x/x_L)\right]
\end{eqnarray}
Where, $x_L$ is the channel periodicity with cross-section maximum and minimum , $y_L$ and $\Delta$, respectively. The cavity and the pore structures are controlled by the exponent $\eta$ [see Fig.~4(c) (inset)]. For $\eta \rightarrow 0$ the channel compartment becomes a 2D rectangular box. For $\eta = 1/2$ the exit window assumes cusp-like strictures as the pores are delimited between two circles tangent to their vertical axis. The corresponding compartment is structurally close to a circular cavity. For $\eta = 1$, the pore tips take triangular shape. For $\eta = 2$, the channel cross-section varies sinusoidally. For large $\eta $ the pore assumes capillary shapes, while the compartments shrink to form a pair of opposite thin dead-ends centred around their midpoint. 

\begin{figure}[H]
\centering \includegraphics[width=7.5cm]{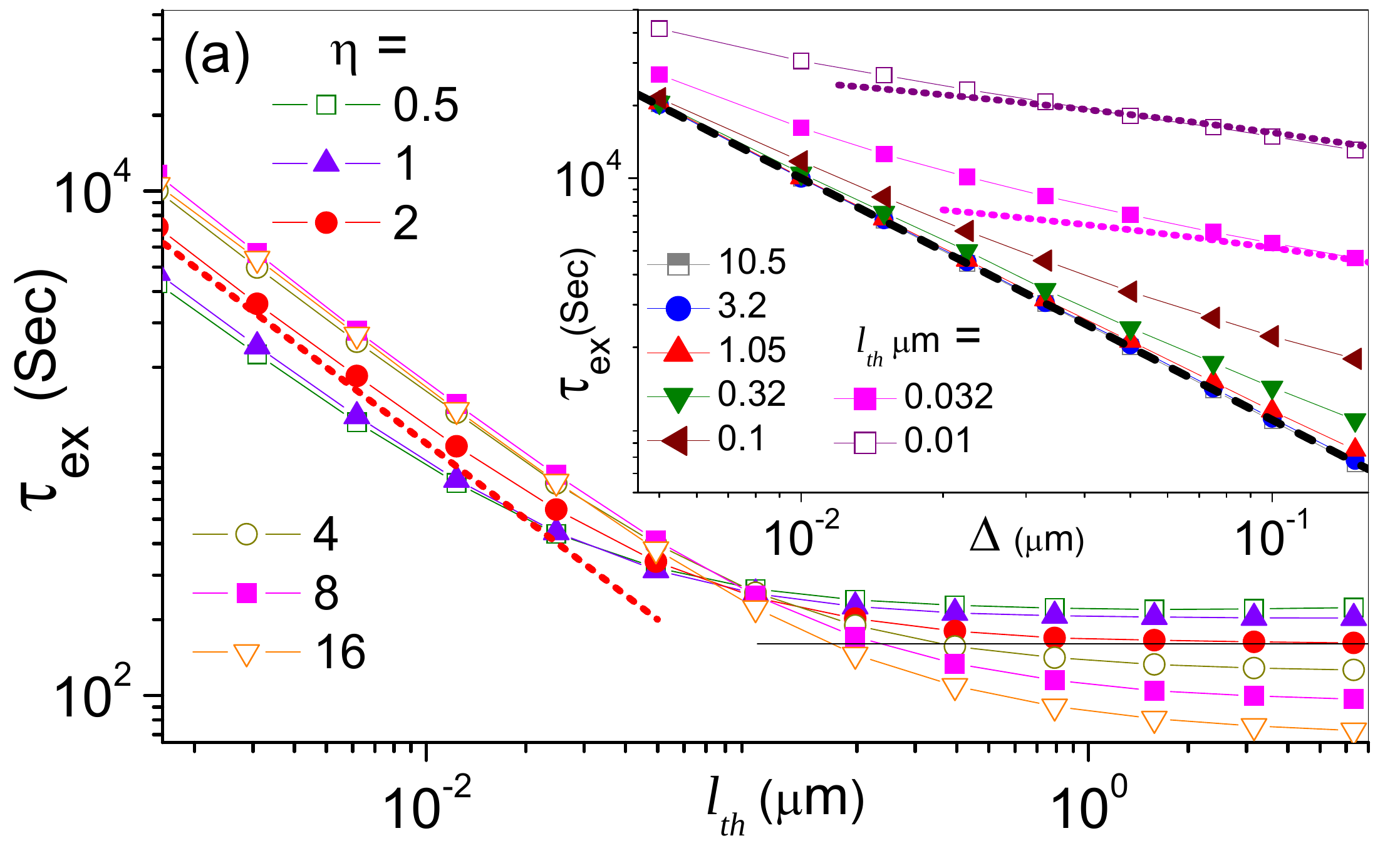}
\centering \includegraphics[width=7.5cm]{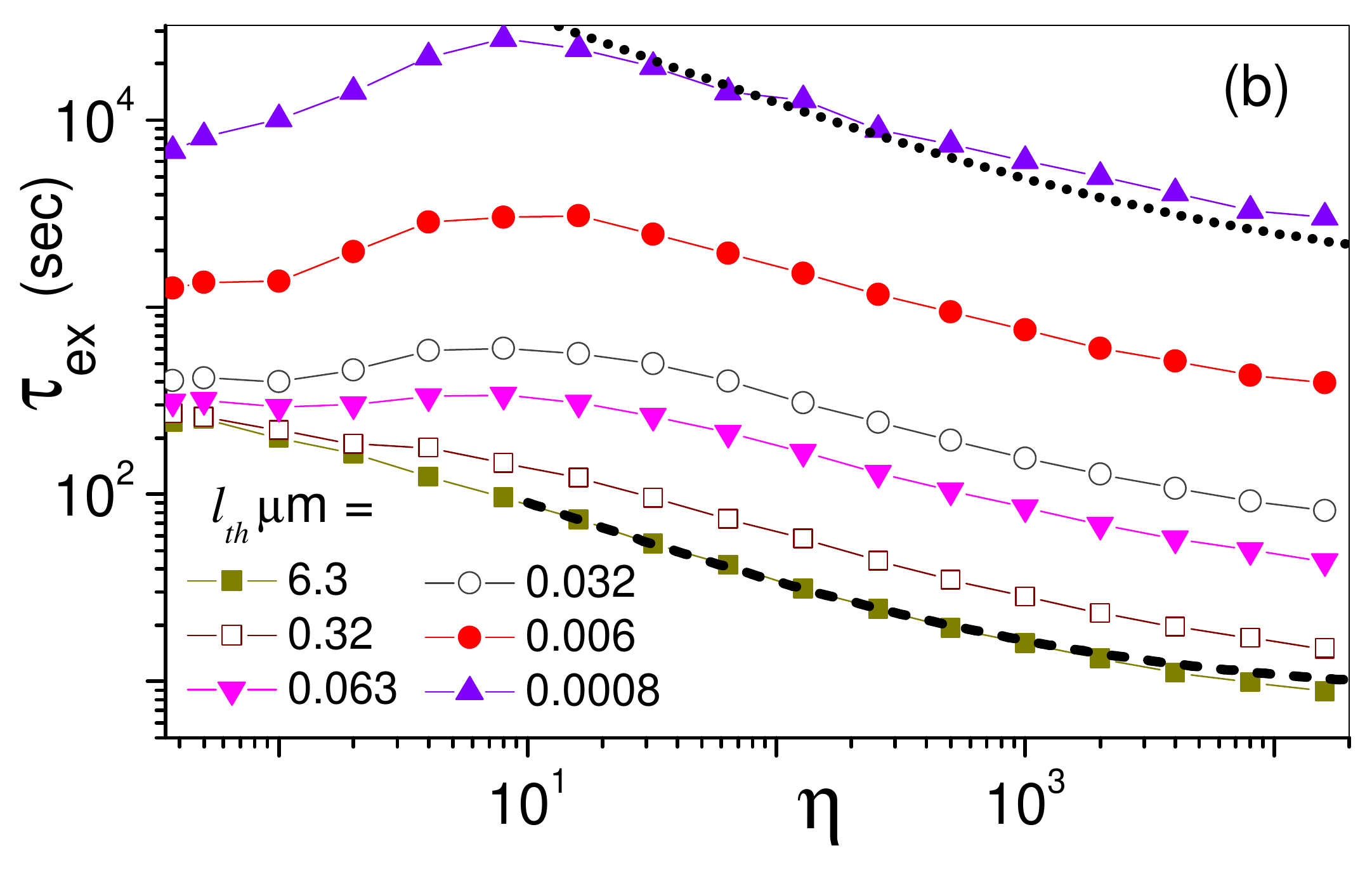}
\centering \includegraphics[width=7.5cm]{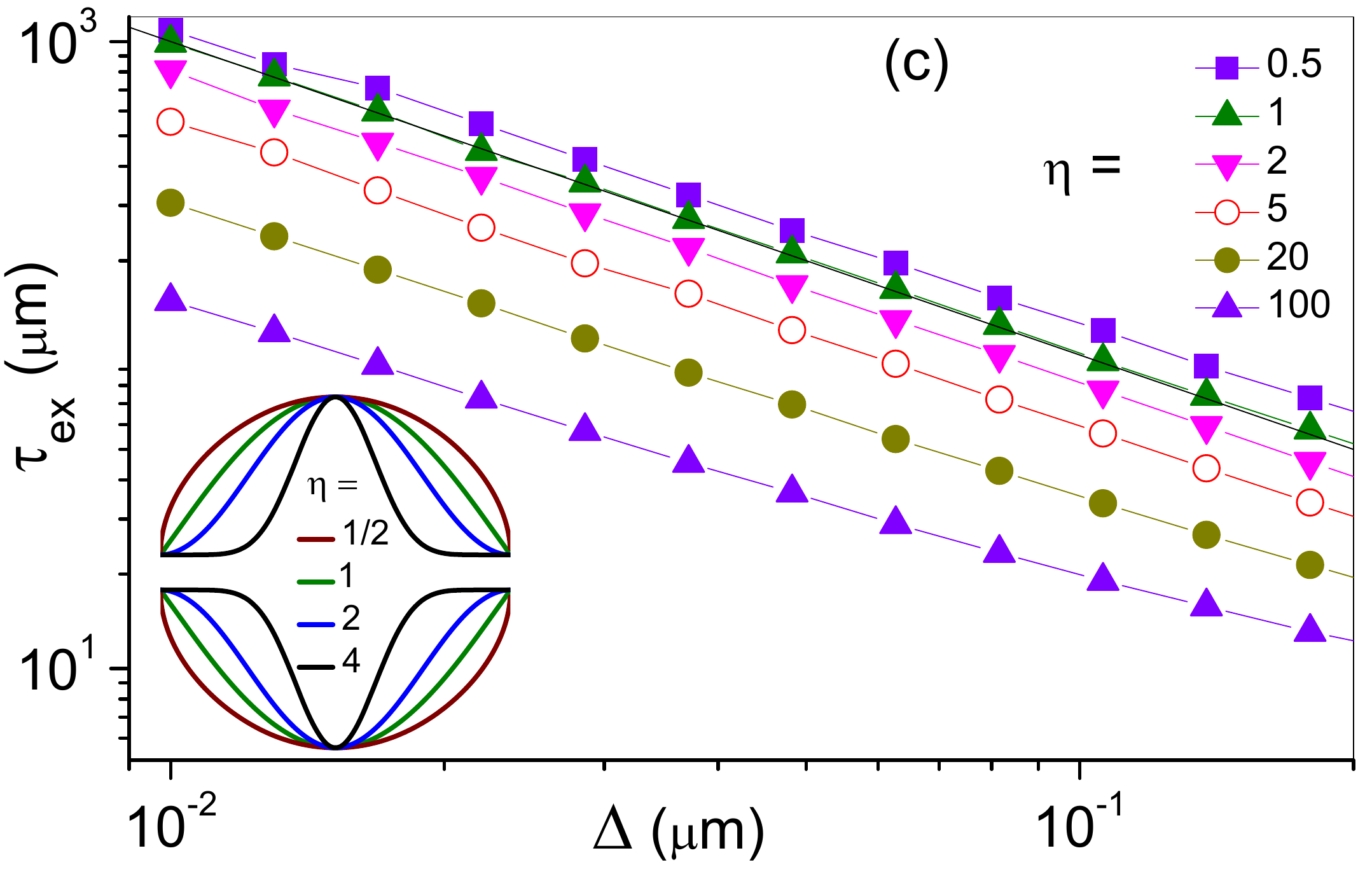}
\caption{(a) $\tau_{ex}$ {\it vs.}  $l_{th} $ (for corrugated channel compartment) with different pore structures (determined by the $\eta$ values). Analytic prediction for sinusoidal corrugated channel in the overdamped limits, $\tau_{ex}=x_L^2/8D_0 \sqrt{y_L/\Delta}$ (see ref\cite{jcp2}), indicated by dotted line. Inset: $\tau_{ex}$ {\it vs.} opening size $\Delta$ (for rectangular cavity) with different $l_{th}$ values. Analytic estimates for low and high damping regimes are represented by dashed line [Eq.~(\ref{rectangular-box-U})] and dotted line [Eq.~(\ref{rectangular-box-O})]. (b) $\tau_{ex}$ {\it vs.} the parameter $\eta$ of the channel
profile function $w(x)$ [see Eq.~(\ref{corrugated-walls})] comparing overdamed and underdamped regimes. The dashed and dotted curves are the relevant analytical predictions based on the Eq.~(\ref{corrugated-2}) for underdamped and overdamped limits, respectively. (c) $\tau_{ex}$ {\it vs.}  $\Delta $ for different $\eta$ and $l_{th} = 6.3\;\mu m$ .  Inset: variation of pore structures with $\eta$. The model parameters are (unless mensioned in the legends): ${\rm v_{th}} = 0.32 \;\mu m/s, \; \Delta = 0.05\;\mu m,\; x_L = y_L = 2\;\mu m. $\label{F4}}
\end{figure}

Let us start with a comparison of the escape times for different structures. Simulation results in Fig.~4(a,b) show that  the exit rate variations with changing pore structures in the overdamped limit are noticeably different from the low damping situation. In the low damping situation, the exit rate first monotonically enhances with increasing $\eta$ and finally reaches to an asymptotic value for $\eta \rightarrow \infty$. On the other hand, in the highly damped condition, the mean exit time first very slowly decays then increases to a maximum with increasing the exponent $\eta$. On further increasing $\eta$ the mean exit time decays to a limiting value as $\eta$ approaches to $\infty$.    

For $\eta =0$, the cavity takes rectangular shape.  In the low damping situation, the exit rate from a rectangular compartment exhibits the similar behaviors as circular cavities. For the thermal length larger than the cavity size ($l_{th}> x_L,\; y_L$), the exit rates are directly proportional to the average velocity and opening size. Based on the same line of reasoning as Eq.~(\ref{MET-UD}), the exit time from the rectangular cavity can be written as,
\begin{eqnarray} \label{rectangular-box-U}
\tau_{ex}= \left(2x_L y_L/\Delta\right) \sqrt{\pi m/2k_B T}
\end{eqnarray}
This analytic estimate well corroborates simulation results [see Fig.~4(a) (inset)]. Upon increasing damping strength beyond a certain value, the deviations from the above estimates become first noticeable  for larger opening size.  For highly damped situation, the exit rate exhibits logarithmic dependence on the pore size~\cite{holcman-review,jcp1,jcp2}, 
\begin{eqnarray} \label{rectangular-box-O}
\tau_{ex}= \left(x_L y_L/2\pi D\right) \ln\left[\left( x_L+y_L\right)/\Delta\right]
\end{eqnarray}
 Estimations based on this equation, depicted in the inset of Fig.~4(a), accord with numerical results for large $\gamma $ (small $l_{th}$) and  $\Delta >>  l_{th}$ . However, simulation results in Fig.~4(a) [inset] deviate from analytic prediction (\ref{rectangular-box-O}) for the very small pore size, as the thermal length becomes comparable to the pore size $\Delta$ and fails to satisfy the condition for overdamped limit, $\Delta >>  l_{th}$.

We now return to detail a major issue, how the escape process from a corrugated channel compartment is affected by the pore structures. Our simulation results show that the exit rate is directly proportional to the opening size irrespective of the pore structures unless $\eta$ is too large. We attribute this result to the fact that the ballistic particle bounces off a compartment wall with a certain rate and only  $\Delta/y_L$ fraction of such collisions leads to pore crossing. Assuming collision frequency is directly proportional to the thermal velocity, the mean exit time can be expressed in a general form,
\begin{eqnarray} \label{corrugated-1}
\tau_{ex}= g(\eta)  \left( x_L \sqrt{\pi m/2k_BT} \right)\left(y_L/\Delta \right)
\end{eqnarray}
The mean exit time is product of three terms: the geometric factor $g(\eta)$, the cell crossing time with a thermal drift velocity $\sqrt{2 k_BT/\pi m}$, and the success probability factor $y_L/\Delta$.

Contrary to the ballistic regime, in the highly damped situation, $\Delta$ dependences of the exit rate guided by the pore geometries ~\cite{jcp3}. For $\eta \le  1$ the exit rate has a logarithmic dependence on the pore width $\Delta$. In case of sinusoidal corrugated channel, $\eta = 2$ the exit rate is directly proportional to $\sqrt{\Delta}$. For $\eta >> 2$ the pore tips grow flatten (takes a capillary like shape) and the exit rate becomes directly proportional to the pore width. 

For very large $\eta$ the exit time $\tau_{ex} \propto \eta^{-1/2} $ and grows insensitive to $\eta$ in the asymptotic limit. Thus, the mean exit time can be expressed as,    
\begin{eqnarray} \label{corrugated-2}
\tau_{ex}= \tau_{\infty}\left(1+  b \;\eta^{-1/2}\right)
\end{eqnarray}
Where, $\tau_{\infty}$ is the mean exit time from the center of a tube of width $\Delta$  and length $x_L$. For $l_{th} >> \Delta$, the particle moves through sequences of correlated bounces extending over the relaxation time $\tau_\gamma$. If the particle is placed at the middle of the tube with a velocity $\bar{v}$ directed at an angle $\theta$ with respect to the channel axis, the exit time out of the tube, $\tau_\theta = x_L \sec{\theta} /2\bar{v}$. However, the angle is restricted between $0$ to $\theta_c$. Where, $\theta_c =\sec^{-1} {(2\bar{v}/\gamma)}$ with an additional restriction $2\bar{v}/\gamma \ge 1$. When $\theta $ exceeds the critical value $\theta_c$, the particle exits the confinement due to velocity relaxation. Thus, the exit time can be assumed  equal to the relaxation time $\tau_{r}$.  Taking average over all possible angles with appropriate weight factor,
\begin{eqnarray} \label{corrugated-3}
\tau_{\infty}= \frac{2}{\pi}\left[\left(\frac{\pi}{2}-\theta_c \right) \frac{m}{\gamma}+ \theta_c \langle \tau_{\theta}\rangle \right]
\end{eqnarray}
Where, $\langle ...\rangle $ is the averaging over the angle $\theta$. It assumes all the angles between $0$ to $\theta_c$ are equally probable. On the other hand, in the highly damped situation, $\tau_{\infty}=\gamma x_L^2/8k_BT$ ~\cite{jcp3}, it is the time to diffuse over the length $x_L/2$. 

The constant $b$ in Eq.~(\ref{corrugated-2}) is a measure of weight factor for contribution from the dead end located at the middle point of the cavity. As mentioned earlier, for larger $\eta$ the compartment cavity shrinks to a central smooth lane of area $\Delta \times  x_L$  and two opposite narrows of cross-section $\Omega_w/x_L$.  Where, $\Omega_w$ is the area encircled by the corrugated walls $w_{\pm}(x)$ can be approximated  as $x_L y_L\sqrt{2/\pi \eta}$ for large $\eta$. Considering the contribution from dead ends and the central channels with statistical weight factors, $\Omega_w/(\Delta + \Omega_w)$ and $\Delta/(\Delta + \Omega_w)$, respectively, we obtain, $b = (x_L /\Delta)\sqrt{2/\pi}$. Note that the cross-section of the central lane is much larger than the width of the dead-end. Further, it is assumed that narrow dead-ends as tubes and the exit time from there is about $2\tau_{\infty}$. Based on this estimation of $b$ and $\tau_{\infty}$, predicted mean exit time is in good agreement with simulation results. It has been shown in Fig.~4(b).

 To this end, key features of escape rate due to variation of cavity structures can be noted as follows. In the ballistic regime, the mean exit time depends on the collision frequency and their success probability. The cavity volume monotonically decreases with increase of $\eta$.  As a result, the collision frequency as well as the exit rate increases  with increase of $\eta$ over the entire range . On the other hand, in the overdamed limit when cavity structures are tuned by the variation of $\eta$ ,  two competing effects, flattening of the pores and shrinking of the compartment area, work on the escape kinetics. Flattening pore structures suppress the exit rate, however, shrinking of cavity volume has  opposite effects.  With increasing $\eta$ the exit time first increases due to flattening of pores. Upon  further increase of $\eta$ beyond a certain value,  the effect of shrinking compartment area dominates. Thus, the mean exit time decreases passing through a maximum.

We conclude this section comparing damping strength dependence of the escape rates for the present analysis with the  energy barrier crossing events. Our simulation results show that all the mean exit time versus damping strength plots have a similar structure irrespective of the pore geometries. The exit time remains insensitive to damping strength until the thermal length $l_{th}$ is larger than the pore width. As we go beyond this limit, the escape time linearly grows with the damping strength. On the other hand, the mean exit time from a meta-stable state (energetic traping) exhibits a minimum in the $\tau_{ex} \; vs. \; \gamma$ which is known as Krammers' turnover~\cite{Hangi-rev2}.

\section{Conclusions}\label{Conclusions}
In this paper we have systematically investigated escape kinetics of a particle from various types of cavities comparing three distinct regimes of damping. In the highly viscous damping, the mean exit rate as well as the other transport quantifiers like diffusivity and mobility are inversely related to the damping strength. However, on decreasing damping below a certain threshold inertia effects manifest itself through deviation of this inverse relation. In the very low viscous medium, when the thermal length grows larger than the cavity size,  particles exhibit ballistic motion within the cavity and occasionally get out of the confinement with a probability proportional to the width of the exit window. Thus, the exit rate becomes directly proportional to the pores width and insensitive to the damping strength. This feature is observed for all the three types of cavities (circular, rectangular and compartments for corrugated channels) and various pore structures. Our simulation results show that  there is quite a large window of viscous relaxation time where diffusion is not ballistic in the length scale of the cavity size, however, observables carry some signature of finite inertia. 

We show that the effects of structural variation of pores on the escape time are considerably different in the two extreme limits of damping. In the highly viscous medium, escape rates enhance on sharpening the pores. However, flattening the pores by increasing $\eta > 2$, the effect of compartment area shrinking dominates, thereby the compartment crossing time decreases with increase of $\eta$.   On the other hand, in the very low damping regime, the escape rate depends on the collision rates and their success probability to get through the exit window. Flattening the pores by increasing $\eta$ reduces the cavity volume and enhances collision frequency on the wall.  Thus, the exit rate enhances with flattening pores for the entire range of $\eta$. 

  All simulation results presented in this paper assume  the particle is placed at the center of the cavity with Maxwellian distribution of velocity at $t =0$. However, the  mean exit time remains  almost unchanged upon changing the initial positions  to uniform  random locations within the cavity [see  Figure S4 in supporting information, page 6-7]. 

Our study shows that in the low damping regimes entropic barrier crossing dynamics exhibits notable different features in comparison to the corresponding energetic cases. Escape rate through an entropic bottleneck is initially insensitive to damping strength. On the other hand, the energy barrier crossing rate  exhibits a turnover behavior with damping strength\cite{Hangi-rev2}, i.e., the escape rate initially increases, however, finally decreases passing through a maximum upon increasing the damping strength.

Our simulation results can be exploited to estimate diffusivity and mobility when applied driving force $F \rightarrow 0 $. Unbiased diffusion through a tube with varying cross-section occurs through sequences of localized motion inside the channel compartments and statistically independent inter-compartment escape events. Such pore crossing events of Brownian particles can be considered as a random walk in 1D with a step length $x_L$ (compartment size) and the time constant $2\tau_{ex}$. Note that the pore crossing time is twice the mean exit time $\tau_{ex}$. Thus, unbiased diffusion through the channel 
formed by joining cavities [see supporting information, Figure S5] can be estimated as, $D(0)=x_L^2/4\tau_{ex}$. Again, at the zero-drive limit ($F\rightarrow 0$), the mobility $\mu$ is related to the diffusivity by Einstein's formula, $\mu = D(0)/D_0$. This relation is restricted to the equilibrium situations. Here, $D_0 = k_BT/\gamma$ is the unbiased diffusion in a free space. Thus, our present analysis could be potentially important to understand transport features of colloidal particles in a variety of corrugated channels. Moreover, dynamics in the underdamped regime is similar to the corrected active Brownian motion. Thus, our results will help realizing transport features of self-propelled particles.

\section*{Supporting Information}
The Supporting Information is available free of charge at
\url{https://pubs.acs.org/doi/10.1021/acs.jpcc.0c04601}.\\
 
\section*{Notes}
The authors declare no competing financial interest.

\section*{Acknowledgments}
D.D. thanks CSIR, New Delhi, India, for support through a
Junior Research Fellowship. P.K.G. is supported by SERB Start-up Research Grant (Young
Scientist) No. YSS/2014/000853 and the UGC-BSR Start-Up Grant No.
F.30-92/2015.

\end{document}